\documentclass[nologo,url,11pt,a4paper]{ETHpaper}
\pdfoutput=1

\usepackage{amsmath, amsthm, amssymb}

\usepackage{graphicx}                  
\usepackage{subfig}                  
\usepackage[numbers,sort&compress]{natbib}

\title{Testing an agent-based model of bacterial cell motility:\\[0.4ex]
How nutrient concentration affects speed distribution}

\author{Victor Garcia, Mirko Birbaumer, Frank Schweitzer}

\address{Chair of Systems Design, ETH  Zurich, Kreuzplatz 5, 8032 Zurich,
Switzerland}
\titlealternative{Testing an agent-based model of bacterial cell motility
\\ \emph{European Physical Journal B} vol. 82, no. 3-4 (2011), pp. 235-244}

\www{\url{http://www.sg.ethz.ch}}

\begin{document}
\maketitle

\begin{center}
\emph{Dedicated to Werner Ebeling on the occasion of his 75th birthday}  
\end{center}

\begin{abstract}
  We revisit a recently proposed agent-based model of active biological
  motion and compare its predictions with own experimental findings for
  the speed distribution of bacterial cells, \emph{Salmonella
    typhimurium}.  Agents move according to a stochastic dynamics and use
  energy stored in an internal depot for metabolism and active motion. We
  discuss different assumptions of how the conversion from internal to
  kinetic energy $d(v)$ may depend on the actual speed, to conclude that
  $d_{2}v^{\xi}$ with either $\xi=2$ or $1<\xi<2$ are promising
  hypotheses. To test these, we compare the model's prediction with the
  speed distribution of bacteria which were obtained in media of
  different nutrient concentration and at different times. We find that
  both hypotheses are in line with the experimental observations, with
  $\xi$ between 1.67 and 2.0. Regarding the influence of a higher
  nutrient concentration, we conclude that the take-up of energy by
  bacterial cells is indeed increased. But this energy is not used to
  increase the speed, with 40$\mu$m/s as the most probable value of the
  speed distribution, but is rather spend on metabolism and growth.
\end{abstract}

\newcommand{\mean}[1]{\left\langle #1 \right\rangle}
\newcommand{\abs}[1]{\left| #1 \right|}
\newcommand{\bbox}[1]{\mbox{\boldmath $#1$}}
\renewcommand{\mathbf}[1]{\bbox{#1}}
\renewcommand{\boldsymbol}[1]{\bbox{#1}}

\section{Introduction}
\label{sec:intro}

Among the many contributions Werner Ebeling made to the interdisciplinary
applications of statistical physics, his concept of \emph{active motion}
stands out as the most proliferate. More than 35 of his own publications
deal, directly or indirecly, with such dynamic phenomena that rely on the
influx of energy. A citation analysis by now mentions 415 citations, lead
by a paper published in \emph{Biosystems} in 1999 \citep{ebeling1999abp}
which also forms the basis of the current publication. But already an
earlier publication in 1994 \citep{steuern-et-94} contained in a nutshell
the main idea of negative friction to accelerate the motion of a Brownian
particle.

The concept of active motion, as proposed by Ebeling, relies on very few,
but reasonable assumptions: particles, which we call \emph{agents} in the
following, have the ability (i) to take up energy from the environment,
(ii) to store it in an internal energy depot, and (iii) to use this
internal energy to accelerate their motion. Without additional energy
take-up, the agent's motion is described by a stochastic dynamics in
terms of a Langevin equation, which denotes the limit case of Brownian
motion. A Brownian particle moves passively because the friction which
would lead to rest, asymptotically, is compensated by a stochastic
force. The fluctuation-dissipation theorem in the special form given by
Einstein tells us how the mean squared displacement of such a Brownian
particle is related to fundamental properties of the medium it is placed
in, such as viscosity or temperature.  This well known scenario is
changed if such particles are turned into ``agents'' by getting
additional internal degrees of freedom, such as the internal energy depot
discussed in the following. Then the passive and random motion can, under
certain conditions, be turned into an \emph{active} and \emph{directed}
motion, which is already found on the level of micro organisms and cells
\citep{birbaumer11:_agent_based_model_intrac_trans}.

The theory developed from the above assumptions makes a number of
predictions about the active motion of agents with an internal energy
depot which have, however, not been tested experimentally. So, it is
worth to find out to what extent living organisms, such as cells or
bacteria, can be described by ``active Brownian particles''. The current
paper wants to contribute to this discussion. In addition to the
theoretical framework already developed, it can build on a parallel
strand of investigations about the motility of cells
\citep{schienbein1993lef}.

The paper is organized as follows: In Sect. 2, we recall the analytical
framework of active Brownian particles, by deriving the equation of
motion in the presence of an internal energy depot. Then, different
assumptions for the conversion of internal into kinetic energy are
developed, which lead to three hypotheses to be tested
experimentally. Sect 3 describes the experimental observations in
detail. A comparison between theory and experiment is carried out in
Sect. 4 at the level of the speed distribution, which is derived from a
Fokker-Planck equation and compared with empirical densities. Details of
the results are presented in the Appendix. A conclusion summarizes our
findings and points out the limitations of their interpretation.

\section{Agent-based model of biological motion}
\label{sec:abm}

\subsection{Internal energy depot}
\label{sec:energy}

Our approach to model the biological motion of bacteria is based on
active Brownian particles or \emph{Brownian agents}
(\citep{schweitzer2003baa}). Each of these agents $i$ is described by
three continuous variables: spatial position $\boldsymbol{r}_{i}$,
velocity $\boldsymbol{v}_{i}$ and internal energy depot $e_{i}$. Whereas
the spatial position and the velocity of an agent describe its
\emph{movement} and can be observed by an external observer, the agent's
energy depot, however, represents an \emph{internal} variable that can
only be deduced indirectly from the agent's motion.

For the internal energy depot we assume, in most general terms, the
following balance equation:
\begin{equation}
  \label{eq:balance}
  \frac{d e_{i}}{d t}= q(\mathbf{r}_{i},t)- w(\mathbf{r}_{i},t)
\end{equation}
$ q(\mathbf{r}_{i},t)$ describes the ``influx'' of energy into the depot,
for example through the take-up of nutrients, which therefore may depend
on the agent's position and on time. The spatially inhomogeneous
distribution of energy was modelled e.g. in
\citep{ebeling1999abp,schweitzer1998cmb}. In the following, we assume
both for simplicity and in accordance with the experiments described
below that nutrients are abundant, hence the take-up of energy is
homogeneous, i.e. constant in time and space, $q=q_{0}(k)$. But,
dependent on the experimental condition, $q_{0}$ varies dependent on the
concentration $k$ of nutrients, but not across agents.

The ``outflux'' of energy from the depot $ w(\mathbf{r}_{i},t)$ depends
on those activities of an agent which require additional energy. In
\citep{schweitzer2002self,ebeling2003self} we have modeled the case that
agents produce chemical information used for communication, e.g. in
chemotaxis. Applying our model to the motion of bacteria, we simply
assume that energy is spent on two ``activities'': (i) Metabolism, which
is assumed to be proportional to the level of internal energy, with the
metabolism rate $c$ being constant in time and equal across
agents. Alternatively, one could assume that metabolism further depends
on the size of the bacteria. (ii) Active motion, i.e. internal energy is
converted into kinetic energy for the bacteria to move at a velocity much
higher than the thermal velocity of Brownian motion. For this conversion
we assume that it proportional to the internal energy and further depends
monotonously, but nonlinearly on the \emph{speed} $v$ of the agent. $v$
is a scalar quantity, describing how fast the agent is moving, regardless
of direction. Velocity $\mathbf{v}$, on the other hand, describes the
direction as well as the speed at which the agent is moving. Our
assumption is that the conversion rate $d(v)$ does not further depend on
the position of the agent or on the direction of motion.
\begin{equation}
  \label{eq:p}
  w(v_{i},t)= e_{i}\left[c + d(v_{i})\right]
\end{equation}
This ansatz satisfies the idea that without internal energy $e$ no
metabolism or active motion is possible.  We note that previously the
particular ansatz $d(v)=d_{2}v^{2}$ was discussed in detail
\citep{ebeling1999abp, condat2002dma} , but not yet confirmed by
experiments. Therefore, in this paper we want to find out whether this or
other possible assumptions are supported by experiments, so we leave
$d(v)$ unspecified for the moment. But it is important to note the
proportion between the two different terms: bacteria spent the vast
amount of their internal energy for metabolism, not for active
motion. Consequently the approximation $c\to 0$, which was discussed in
previous investigations, does not hold for bacteria.

Assuming that the internal energy depot relaxes fast into a
quasi-stationary equilibrium allows to approximate the internal energy
depot as
\begin{equation}
  \label{eq:estat}
  e^{\mathrm{st}}_{i}(v_{i})=\dfrac{q_0}{c+d(v_{i})}  
\end{equation}
I.e., the level of the internal energy depot follows instantaneously
adjustments of the speed.

\subsection{Equation of motion}
\label{sec:motion}

The equation of motion for a Brownian agent is given by a Langevin
equation for the velocity $\mathbf{v}_{i}$. However, because of the
conversion of internal into kinetic energy with a rate $d(v)$, we need to
consider an additional driving force the structure of which can be
obtained from a total energy balance. In the absense of an external
potential, the mechanical energy of the agent is given by the kinetic
energy, $E_{i}=mv_{i}^{2}/2$, which can be changed by two different
processes: (i) it decreases because of friction, with $\gamma$ being the
friction coefficient (equal for all agents), and (ii) it increases
because of conversion of internal energy into energy of motion. Hence,
with $\mathbf{v}_{i}=\dot{\mathbf{r}}_{i}$ and
$\dot{\mathbf{v}}_{i}=\ddot{\mathbf{r}}_{i}$ the total balance balance
equation gives:
\begin{equation}
  \label{eq:ebal}
  \frac{d}{dt}E_{i} = 
  m \dot{\mathbf{r}}_{i} \ddot{\mathbf{r}}_{i}= 
  e d(v_{i})-m \gamma  v_{i}^2
\end{equation}
Deviding this equation by $m\dot{\mathbf{r}}_{i}$ results in the
deterministic equation of motion:
\begin{equation}
  \label{eq:det}
  \dot{\boldsymbol{v}}_{i} 
  = - \boldsymbol{v}_{i} \left[\gamma
    - \frac{e}{m}\frac{d(v_{i})}{v_{i}^2} \right]
\end{equation}
Based on this, we propose the Langevin equation for the Brownian agent by
adding to the right-hand side of eqn. (\ref{eq:det}) a stochastic force
$\boldsymbol{\mathcal{F}_{i}}(t)$ with the usual properties, namely that
no drift is exerted on average, $\mean{\mathbf{\mathcal{F}}_{i}(t)} = 0$,
and that no correlations exist in time or between agents,
$\mean{\mathbf{\mathcal{F}}_{i}(t) \mathbf{\mathcal{F}}_{j}(t')} = 2
S~\delta_{ij}\delta(t-t')$. For physical systems the strength of the
stochastic force $S$ is defined by the fluctuation-dissipation theorem
which itself builds on the equipartition law, but for microbiological
objects such as cells and bacteria the situation has proven to be more
intricate, as we will outline later.

Assuming a quasistationary internal energy depot, eqn. (\ref{eq:estat})
and defining $\gamma_{0}=\gamma m$, 
we arrive at the modified Langevin equation for the Brownian agent:
\begin{equation}
  \label{eqno3}
  \dot{\boldsymbol{v}}_{i} = - \gamma \boldsymbol{v}_{i} 
  \left[1  - \frac{q_{0}}{\gamma_{0}} \dfrac{d(v_{i})}
    {\left \lbrack c+d(v_{i})
      \right \rbrack {v}_{i}^2}
  \right]  + 
  \sqrt{2S}\ \boldsymbol{\xi}_{i}(t)
\end{equation}
where $\mathbf{\xi_{i}}$ denotes Gaussian white noise. Eqn. (\ref{eqno3})
shall be used as the starting point for the further discussion.

\subsection{Hypotheses for $d(v)$}
\label{sec:hyp}

We now specify the function $d(v)$ for which we assume a nonlinear
dependence on the speed in terms of the following power series:
\begin{equation}
  \label{powerseries}
  d(v)=\sum_{k=0}^{n}d_{k} {v}^{k}.
\end{equation}
Using different orders of this power series, we first evaluate the
stationary velocity estimated from the deterministic part of
eqn. (\ref{eqno3}) (omitting the index $i$ for the moment) and then
compare the outcome against data from experiments with bacteria.

Neglecting the stochastic term of eqn. (\ref{eqno3}), we first notice
that $\bbox{v}=0$ is always a solution. Caused by friction, the motion of
an agent shall come to rest -- but in a stochastic system agents still
passively move thanks to the impact of the random force
$\mathbf{\mathcal{F}}$.

Secondly, taking into account the influence of the internal energy depot,
we notice the importance of the rate of metabolism, $c$. For $c=0$, we
always find for the nontrivial speed
\begin{equation}
  \label{eq:c0}
  v_{s} =  \sqrt{\frac{q_{0}}{\gamma_{0}}}\;; \quad (c=0,\;\gamma_{0}=\gamma m)
\end{equation}
independent of further assumptions for $d(v)$. I.e., dependent on the
take-up of energy $q_{0}$ agents move with a constant speed. For $c\neq
0$ and $n=0$, i.e. for $d(v)=d_{0}$, this nontrivial speed is corrected
leading to
\begin{equation}
  \label{eq:d0}
  v_{s} = \left(\frac{q_{0}}{\gamma_{0}}\right)^{1/2}\, 
  \left(\frac{1}{1+(c/d_{0})}\right)^{1/2}
\end{equation}
That means dependent on the proportion of metabolism vs active motion,
the stationary velocity can be consirably lowered. For further comparison
it is convenient to rewrite eqn. (\ref{eq:d0}) in a different way
\begin{equation}
  \label{eq:c0d0}
  v_{s} =\left(\frac{c}{d_{0}}\right)^{1/2}\, 
  \left(\frac{Q_{0}}{1+(c/d_{0})}\right)^{1/2} \;; \quad 
  Q_{n}=\frac{q_{0}d_{n}}{\gamma_{0}c} \;; \quad (n=0)
\end{equation}
where the reduced parameter $Q_{n}$ combines the relevant parameters of
the model describing take-up energy and active motion $(q_{0},d_{n})$ and
external and internal dissipation $(\gamma_{0},c)$.

Considering the next higher order of the polynom, $d(v)=d_{1}v$ for
$n=1$, the stationary speed follows from the quadratic equation:
\begin{equation}
  \label{n12}
  v_{s}^{2}+\left (\frac{c}{d_{1}}\right ) v_{s} -\frac{q_{0}}{\gamma_{0}}=0
\end{equation}
One can verify that these solutions are not consistent with other
physical considerations, in particular the speed $v_{s}$, for large $c$, is
biased toward negative values. Testing another first order assumption,
$d(v)=d_{0}+d_{1}v$, does not improve the situation, because $d_{0}$ is
additive with $c$ and thus just rescales the metabolism rate.

Consequently, we have to restrict ourselves to the case $n=2$, i.e. we
arrive at $d(v)=d_{2}v^{2}$ which is the known SET model
\citep{schweitzer1998cmb,condat2002dma} with the stationary solution for
the speed $v$:
\begin{equation}
  \label{eq:set}
  v_{s}=\sqrt{\frac{q_{0}}{\gamma_{0}}-\frac{c}{d_{2}}}
  =
  \left(\frac{q_{0}}{\gamma_{0}}\right)^{1/2}\left(1-\frac{1}{Q_{2}}\right)^{1/2}
  = 
  \left(\frac{c}{d_{2}}\right)^{1/2}\left(Q_{2}-1\right)^{1/2}
\end{equation}
Different from the previous cases, for $n=2$ we find a bifurcation
dependent on the control parameter $Q_{2}$. For $Q_{2}\leq 1$, $v=0$ is
the only real stationary solution, whereas for $Q_{2}>1$ a nontrivial
solution for the speed exist. The possible consequences are already
discussed in the literature. In \citep{erdmann2000bpf, schweitzer2001smc}
the supercritical case, $Q_{2}>1$ was investigated, while in
\citep{condat2002nem, condat2002dma, sibona2007eml} the subcritical case
$Q_{2}\leq 1$ was considered.  Because metabolism consumes the lion share
of the internal energy provided, one would assume that $Q_{2}\ll 1$ is
the most realistic case for the motion of bacterial cells -- which
remains to be tested.

For the stochastic motion, eqn. (\ref{eqno3}), in the supercritical case
the contribution of the stochastic term is small compared to the kinetic
energy provided by the energy depot. Hence, the agent should move forward
with a non-trivial velocity (i.e. much above the thermal fluctuations),
which has a rather constant speed, but can change its direction
occasionally. In the subcritical case, on the other hand, the stochastic
fluctuations dominate the motion, but the energy depot still contributes,
this way resulting in the first order approximation of the stationary
velocity (in one dimension): $v^{2}= k_{B}T/m(1-Q_{2})$
\citep{condat2002nem}.  In fact, the authors of \citep{condat2002nem} put
forward a nice argument that in the high dissipation regime -- or in
environments with low nutrition concentration -- a strong coupling
between the two energy sources (depot and noise) appears that should help
micro organisms to search more efficiently for a more favorable
environment.

We are not going to repeat these theoretical discussions. Instead, we ask
a different question not investigated so far: which of the above cases is
consistent with \emph{experimental findings}? As outlined above, the SET
model with its two regimes, (i) $Q_{2}\leq 1$, i.e.  subcritical energy
supply, and (ii) $Q_{2}>1$, i.e. supercritical energy supply, is the most
promising ansatz to be tested for $d(v)$. To compare this with a more
general setting, instead of integers $n=0,2$ we may also consider
fractional numbers $n=\xi$ with $0<\xi<2$, i.e. $d(v)=d_{2}{v}^{\xi}$,
which results in the following equation for the stationary solutions:
\begin{equation}
  \label{eq:xi}
  v^{\xi}-\left(\frac{q_{0}}{\gamma_{0}}\right)v^{\xi-2}+\frac{c}{d_{2}}=0
\end{equation}
Reasonable values of $\xi$ should be in the interval between 1 and 2 --
for which we expect two nontrivial solutions for the stationary velocity,
but no bifurcation with respect to the parameter $Q_{2}$. In conclusion,
our theoretical investigations provide us with three different hypotheses
for the active motion of biological agents:
\begin{enumerate}
\item $d(v)=d_{2}v^{2}$ with subcritical take-up of energy,
  i.e. $Q_{2}<1$
\item $d(v)=d_{2}v^{2}$ with supercritical take-up of energy,
  i.e. $Q_{2}>1$
\item $d(v)=d_{2} {v}^{\xi}$ with $1\leq \xi \leq 2$ and $Q_{2}>1$
\end{enumerate}

\section{Experimental Observations}
\label{sec:exp}

\subsection{Experimental Setup}
\label{sec:setup}

In order to test which of the above outlined hypotheses regarding active
biological motion is compatible with real biological motion of bacteria,
we proceed as follows: Bacterial cells are placed in a shallow medium
that can be approximated as a two-dimensional system. Keeping all other
conditions constant, we vary the nutrition concentration in that medium
so that three different nutrition levels are maintained: low, medium, and
high. We then measure the velocity distribution of the bacteria (as
described below) in response to these nutrition levels.
Eventually, we do a maximum likelihood estimation of the parameters
describing the velocity distributions and compare these with the
hypotheses made. Ideally, we would expect that the experimental velocity
distributions could be better fitted by one of the hypotheses, while the
others could be rejected.

When it comes to the specific setup of the experiments, we soon realize
that the devil is in the details. In order to be conform with our
hypotheses, we would need to test bacteria that move like Brownian
particles in the limit of low nutrition, while performing a rather
directed motion for high nutrition concentration, with arbitrary changes
in the direction.  Instead, most bacteria, \emph{Escherichia coli} being
a prominent example, move quite differently, i.e. their movement switches
between tumbling and nontumbling phases \citep{condat2005rcr}.  Tumbles
denote temporary erratic movements, whereas during the nontumbling
phases, called runs, bacteria execute a highly directed, ballistic-like
motion. Both of these phases describe a different form of active motion,
but do not differ in the mechanism or level of energy supply. Precisely,
the flagellar propellers responsible for the forward motion
\citep{berg2003rmb, magariyama2001bss, berry1999bfm, manson1980efr}
rotate with the same efficiency during the tumbling and non-tumbling
phases \citep{chen2000tsr}.

In order to avoid an abritary averaging over these different forms of
active motion, we have chosen to study bacteria that do not tumble at
all, specifically the non-tumbling strain M935 of \textit{Salmonella
  typhimurium} \citep{stecher2008mas,winnen2008hep}. This type of
bacterial cells has another advantage in that it does not perform
chemotaxis, i.e. it does not follow chemical gradients or gradients in
the nutrient concentration, which would bias the motility towards
directed motion. But mutant strain is capable to take up the nutrients at
the same rate as normal \textit{S. typhimurium}.

For the medium, we have realized an almost two-dimensional setup, keeping
in mind that three-dimensional motion results in a projection error of
the trajectories.  Further, we need to ensure that both the temperature
$T$ and the viscosity $\eta$ of the medium is kept constant over time and
across setups with different nutrient concentrations. These were prepared
as follows:
\begin{description}
\item[Medium 0] Used as a reference case where no additional nutrients
  are available for the bacteria. It consists of a
  \emph{phosphate-buffered saline working solution} (\textbf{PBS}), with
  5 \
  protein.
\item[Medium 2] A nutritionally rich medium which contains of
  \emph{lysogeny broth} (\textbf{LB}), a substance also primarily used
  for the growth of bacteria.
\item[Medium 1] A medium with an intermediate concentration of
  nutrients. It contains an mixture of medium 0 and 2, equally.
\end{description}
We assume that the nutrients are equally dissolved in the whole medium
and that nutrient concentration differences can be neglected. Further,
the depletion of nutrients due to consumption of the bacterial cells can
be neglected. Hence, we assume that the take-up of energy per time unit,
$q_{0}$ is constant and equal for all bacterial cells, but may change
dependent on the nutient concentration, i.e.  $q_{0}$ needs to be
determined for each of the three different settings as described below.

\subsection{Measuring trajectories and velocities}
\label{sec:traj}

In order to observe trajectories, bacteria of the same \emph{Salmonella}
strain were grown (details see Appendix A) and were put into the three
different media. For each medium, we recorded two movies at different
times: (i) after the bacteria were put into the different media (initial
condition), and (ii) after about one hour, i.e. after a sufficiently long
time of relaxation, which ensures a stationary velocity distribution.  We
had to assume that the cells would adapt rather slowly to the new
environment. Appendix B presents the details of how the trajectories were
recorded.

\begin{figure}[htbp]
  \centering
  \includegraphics[height=7.5 cm,
  angle=-90]{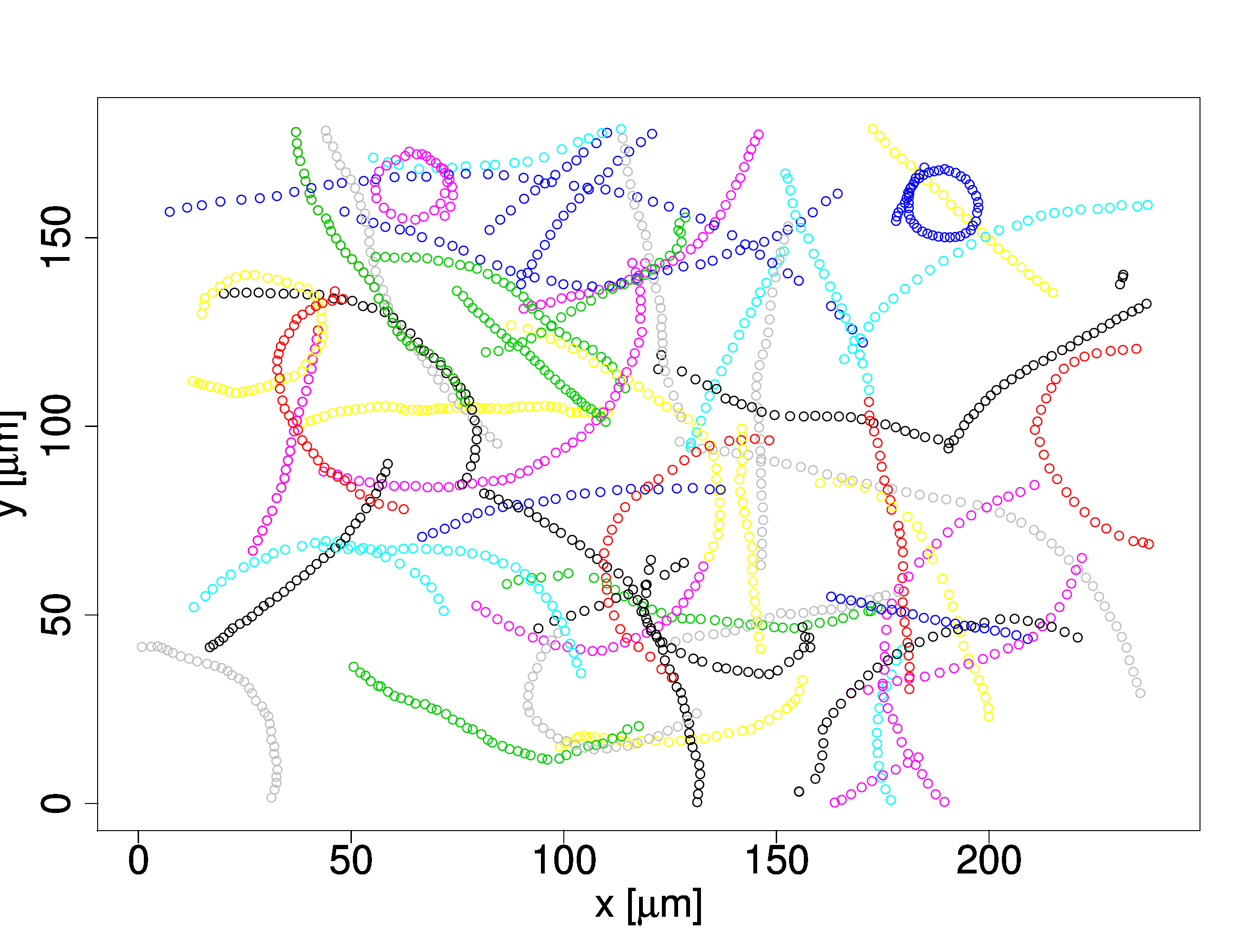}\hfill
  \includegraphics[height=7.5cm,
  angle=-90]{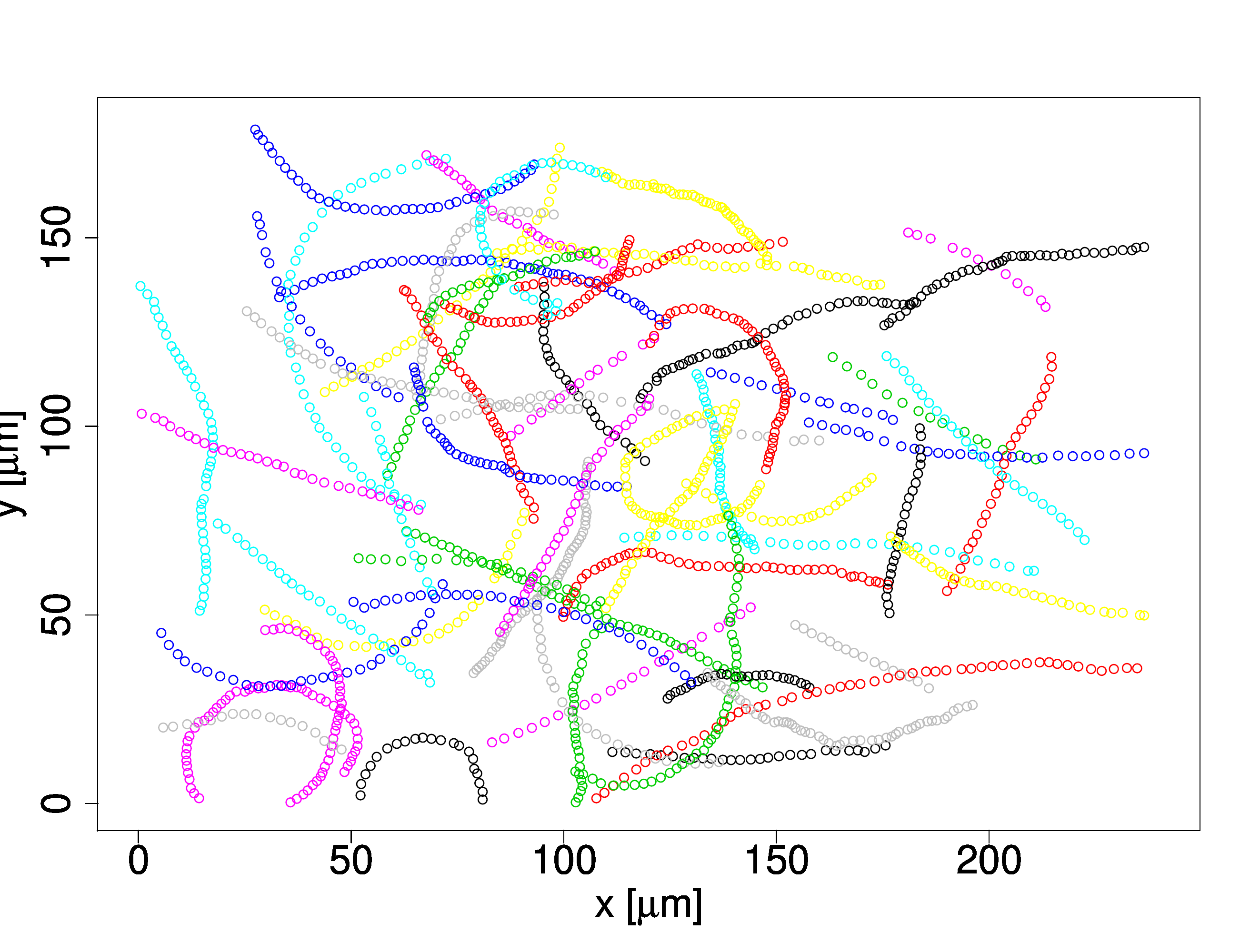}
  \caption{All bacterial trajectories recorded in medium 0 (no nutrients)
    at times $t_{0}$=0min in (top, 54 trajectories) and
    $t_{\mathrm{end}}$=84min (bottom, 61 trajectories). At both times,
    the movie length evaluated was about 17 sec with a time resolution of
    0.11 sec, i.e. about 700 frames per movie. The points of the
    trajectories show 26 of these frames.  The rectangular boundaries are
    a consequence of the camera calibration.  }
  \label{AlltrajPBS4-S7}
\end{figure}

As the trajectories of Fig. \ref{AlltrajPBS4-S7} verify for both the
initial and the stationary conditions, \emph{Salmonella typhimurium} swim
in quasi-ballistic manner through the fluid. Their trajectories are
mostly slightly curved. Initially, at least two trajectories are very
strongly curved where bacteria seem to swim in narrow circles.  The
videos make clear that bacteria swim in an isotropic manner and with a
mean velocity. It is remarkable that the bacteria maintain their
quasi-ballistic movement even after more than one hour, despite not being
able to take up energy from the medium.

\begin{figure}[htbp]
  \centering
  \includegraphics[width=0.3\textwidth,angle=0]{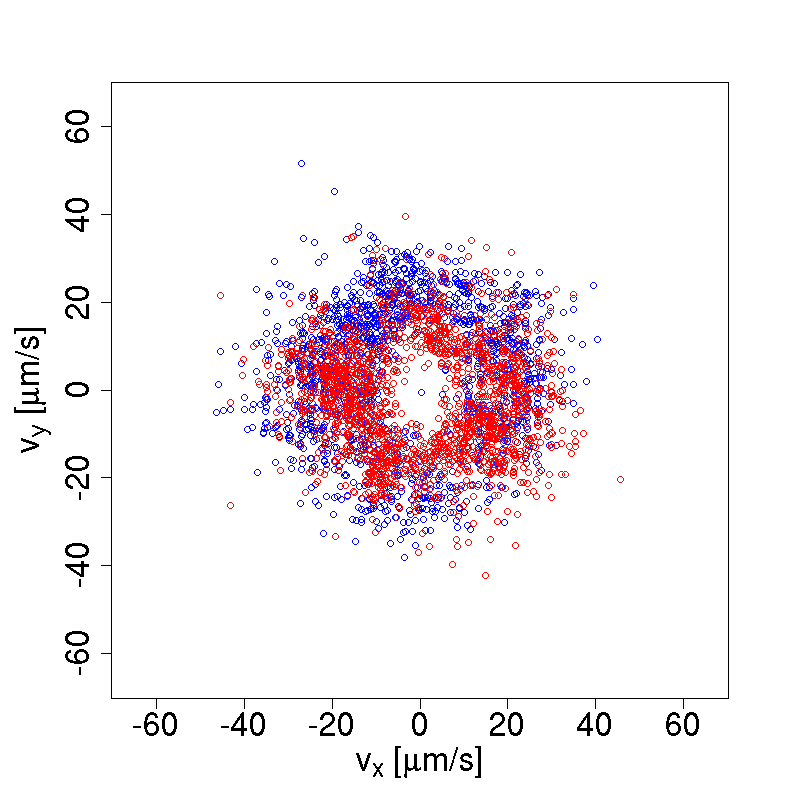}\hfill
  \includegraphics[width=0.3\textwidth,angle=0]{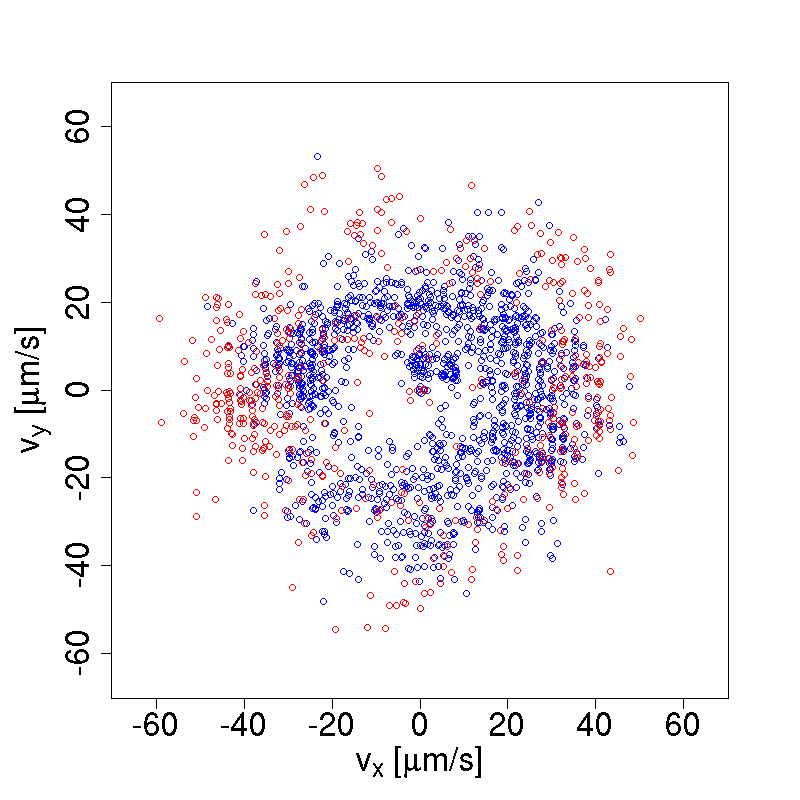}\hfill
  \includegraphics[width=0.3\textwidth,angle=0]{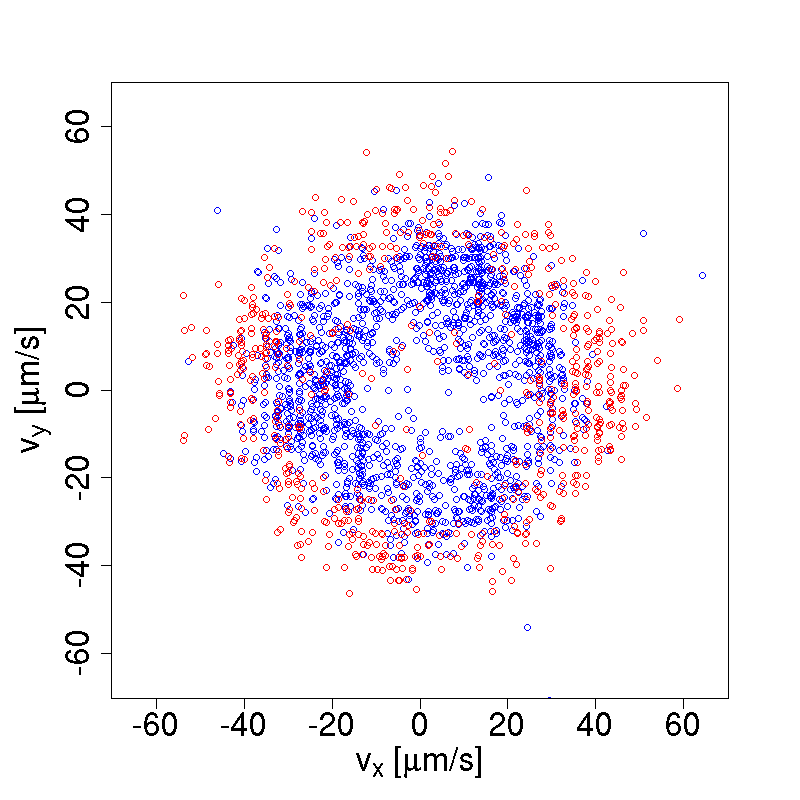}
  \caption{Snapshots of the two-dimensional velocity distributions around
    times $t_{0}$ (blue dots) and $t_{\mathrm{end}}$ (red dots), shown
    for the three different media: (left) medium 0, (middle) medium 1,
    (right) medium 2. Each of the samples contains about 1.500 data
    points.}
  \label{aveldis}
\end{figure}

Given the trajectories measured at a time resolution of 0.11 sec, we are
able to calculate the velocity vector in the two-dimensional space as
\begin{equation}
  \label{calc}
  \boldsymbol{v}^{k}_{t_j}=\frac{\boldsymbol{r}^{k}_{t_j}
    -\boldsymbol{r}^{k}_{t_{j-1}}}{t_j-t_{j-1}}
\end{equation}
where $k$ denotes the bacterial cell and $j$ refers to the time step.
The velocity distributions are shown in Fig. \ref{aveldis} both for the
initial situation (blue dots at $t_{0}$) and after a sufficient long time
of relaxation (red dots at $t_{\mathrm{end}}$) and for the three
different media used.  Each of the samples contains about 1.500 data
points.

In all cases, we observe the formation of rings of different radii,
indicating that bacteria swim at comparable velocities in the time
intervals of observation and that their motion is isotropic, i.e. that
they do not have a preferrred direction of motion. Most interesting,
compared to the initial distribution the rings either contract (medium 0)
or expand (medium 1, 2) in diameter, which means that the bacteria have
adjusted their individual velocities according to the nutrient
concentration available. This will be systematically investigated in the
next section. The available data do not allow us to predict if the rings
completely contract (medium 0) or further expand (medium 1,2 ) their
extension.

\section{Investigating the velocity distribution}
\label{Results}
\subsection{The Fokker-Planck Perspective}
\label{sec-FPp}

The experiments described above have clearly shown that bacteria adjust
their velocity dependent on the nutrient available in the medium. It
remains (i) to quantify this influence, and (ii) to compare the outcome
with the hypothesis made on the velocity dependent transfer of internal
energy. Such a comparison cannot be made on the level of individual
trajectories, but only on the level of the ensemble average.

Hence, in the following we use the two-dimensional velocity distribution
$p(\mathbf{v},t)$, which follows a Fokker-Planck equation
\citep{erdmann2000bpf, schweitzer2001smc} that corresponds to the
Langevin eqn. \ref{eqno3}):
\begin{equation}
  \label{FokkPlSchweitzer}
  \frac{\partial p(\boldsymbol{v},t)}{\partial t}= 
  \boldsymbol{\nabla}_{v}  
  \left \lbrack \gamma \mathbf{v} 
    \left (1 - \frac{q_0}{\gamma_{0}} \frac{d(v)}{\left[c+d(v)\right] \boldsymbol{v}^2} \right )
    \boldsymbol{v}\; p(\boldsymbol{v},t)  
    + S~\boldsymbol{\nabla}_{v} p(\boldsymbol{v},t)  \right \rbrack
\end{equation}
This Fokker-Planck equation is based on the assumption that the internal
energy depot has reached a quasistationary equilibrium fast enough. If we
assume the general hypothesis $d(v)=d_{2}{v}^{\xi}$ and $\dot{p}=0$, we
find for the stationary velocity distribution
\begin{equation}
  \label{pv0}
  p^0(\boldsymbol{v})= C \left( 1+ \frac{d_2 {v}^\xi}{c} 
  \right)^{\frac{q_0}{\xi m S}}
  \exp \left( - \frac{\gamma}{2 S} \boldsymbol{v}^2 \right ),
\end{equation}
where the normalization condition $C$ is defined through the condition
$1=\iint p^0(\boldsymbol{v}) d\boldsymbol{v}$. For $\xi=2$ the SET model
results and the stationary solution, eqn. (\ref{pv0}), can be written in
first-order expansion as:
\begin{equation}
  \label{pv0-exp}
  p^0(\boldsymbol{v}) \sim 
  \exp \left( - \frac{\gamma}{2 S} \left[ 1- Q_{2}\right]  \boldsymbol{v}^2
    + \cdots \right)
\end{equation}
As has been discussed in detail \citep{erdmann2000bpf}, for $Q_{2}<1$ we
find a unimodal Maxwell-like velocity distribution, whereas for $Q_{2}>1$
a crater-like velocity distribution results in two-dimensional systems.

\begin{figure}[htbp]
  \centering
  \includegraphics[width=7cm] {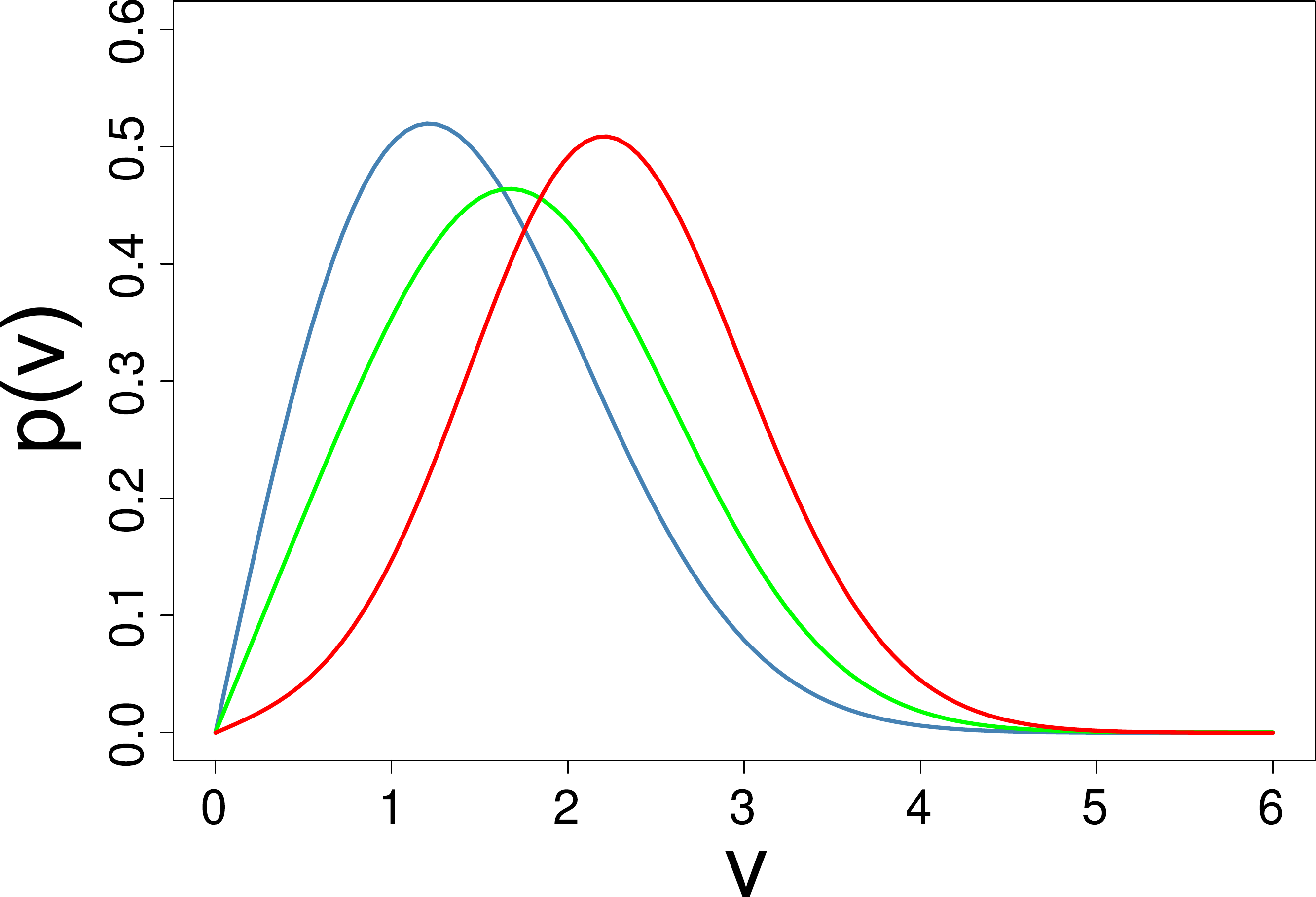}
  \caption{$p_a^0(v)$, eqn. (\ref{pa0}), for $\xi=2$ and different values
    of $Q_{2}$: (blue) $Q_{2}=0.35$, (green) $Q_{2}=1$, (red)
    $Q_{2}=3.5$. The speed $v$ is given in abitrary units.}
  \label{VariationD2}
\end{figure}

The amount of data measured does not allow us to reasonably reconstruct
the two-dimensional velocity distribution by means of density
approximations. Therefore, in the following, we restrict ourselves to the
distribution of the speed which contains sufficient information to test
our hypotheses given.  To find the speed distribution $p^0_a(v)$ in two
dimensions we integrate over a disk $B(v')$:
\begin{equation}
  P^0(|\boldsymbol{v}|<v')= \iint_{B(v')} p^0(\boldsymbol{v}) dv_x dv_y= 
  2\pi \int_0^{v'}p^0(v)dv= \int_0^{v'}p^0_a(v)dv
\end{equation}
and find
\begin{equation}
  \label{pa0}
  p^0_a(v)=C_{a} 2 \pi  v \left( 1+ \frac{d_2 {v}^\xi}{c}
  \right)^{\frac{q_0}{\xi m S}}
  \exp \left( - \frac{\gamma}{2S} v^2 \right )
\end{equation}
When comparing this equation with eqn. \ref{pv0}, one should note the
additional prefactor $v$. So, for the maximum of $p_{a}^{0}(v)$, instead
of the compact expression (\ref{eq:set}) resulting from
${p}^{0}(\mathbf{v})$, we find a rather intricate expression which is not
reprinted here. Instead, the stationary speed distribution is plotted in
Fig. \ref{VariationD2} for the SET model and different values of
$Q_{2}$. Different from the noticable change between unimodal and
crater-like shape of the stationary velocity distribution
$p^{0}(\mathbf{v})$, we observe only a shift in the maximum of
$p_{a}^{0}(v)$ when $Q_{2}$ changes from subcritical to supercritical
values. At second view, one notices that the \emph{ascent} of
$p_{a}^{0}(v)$ changes from a linear increase for $Q_{2}<1$ to an
nonlinear increase $v^{n}$ for $Q_{2}>1$, where $n$ is an integer.

Specifically, we do not need to explicitely derive the maximum of the
speed distribution (which is also known as the most probable value and
different from, e.g., the expecation value or the average value), because
we want to compare the theoretical and the experimental
\emph{distributions} $p_{a}(v)$ rather than their extreme values. These
distributions, in addition to their mean value, are further characterized
by their width, given by the variance $\sigma^{2}=S/\gamma$. Hence, we
need to determine the strength $S$ of the stochastic force in relation to
the friction coefficient $\gamma$.

\subsection{Determining the noise intensity}

In statistical physics, the strength of the stochastic force is related
to the thermal velocity of microscopic particles, e.g. molecules or
Brownian particles, via the fluctuation-dissipation theorem, which yields
for ideal gases $S/\gamma =k_{B}T/m$, where $k_{B}$ is the Boltzmann
constant. If one wishes to apply the same relation also to bacteria like
\emph{Escherichia coli} or \emph{Salmonella typhimurium} at about
$T=300$, with a bacterial mass of $m\approx 10^{-15}$, one arrives at
$\sigma=2000 \mu$m/s, which about two orders of magnitude larger than the
average velocity for such bacteria. Given the rather complex nature of
bacterial motion described above, this discrepancy is not really
surprising.

Therefore, in \citep{schienbein1993lef} a different method to determine $S/\gamma$
was proposed, which is based on the speed autocorrelation function 
\begin{equation}
 g_{v}(\tau)=\frac{\mean{v(t_{1})v(t_{2})}}{\mean{v(t_{1})v(t_{1})}}\;;
\quad \tau=\abs{t_{2}-t_{1}}
\label{eq:autocorr}
\end{equation}
The calculation of $g_{v}(\tau)$ needs a formal solution of the Langevin
equation for the speed ${v}(t)$. This was provided in \citep{schienbein1993lef}
for a different model applied to the migration of human granulocytes. It
postulates a stationary speed $v_{s}$ and assumes, different from our
ansatz of eqn. (\ref{eqno3}),
\begin{equation}
  \label{eq:gruler}
    \dot{{v}}_{i} = - \gamma {v}_{i} 
  \left[1  - \frac{v_{s}}{v}
  \right]  + 
  \sqrt{2S}\ {\xi}_{i}(t)
\end{equation}
It was noted that in this equation an additive Gaussian random noise is
physically and mathematically problematic as it allows in principle
negative speed values. As possible solution, one can consider
appriopriate boundary conditions or a different definition of the
velocity-dependent friction term as suggested recently
\citep{PhysRevLett.106.230601}

In the following, we still use eqn. \ref{eq:gruler}, but point out that
this is only an approximation in the limit of small noise with respect to
the finite stationary speed.  The stationary solution for the speed
autocorrelation function is reached when the times $t_{1}$ and $t_{2}$
are larger than the characteristic time $\gamma^{-{1}}$, which results
into
\begin{equation}
  \label{eq:autogruler}
  \mean{v(t_{1})v(t_{2})}=v_{s}^{2} + 
  \frac{S}{\gamma}\exp\left\{-\gamma|t_{2}-t_{1}|\right\}
\end{equation}
In the limit $\tau \to \infty$, the speed autocorrelation function
$g_{v}(\tau)$ becomes a constant, $g_{v}={v_{s}^{2}}/[{v_{s}^{2}+
  {S}/{\gamma}}]$, which can be measured experimentally, to obtain:
\begin{equation}
  \label{eq:gvconstant}
\sigma^{2}= \frac{S}{\gamma} = v_{s}^{2} \left(\frac{1- g_{v}}{g_{v}}\right)
\end{equation}
\citet{schienbein1993lef} found for human granulocytes, which are much
larger than the bacterial cells investigated here, $g_{v}=0.82$ and from
the measured speed distribution the maximum value
$v_{s}=17\mu$m/min. 

As mentioned, our Langevin equation uses a different ansatz for the
velocity-dependent friction function, which does not allow us to obtain a
simple closed form for $g_{v}$. However, we can still apply the results
of \citep{schienbein1993lef} arguing that the speed of bacteria in the
stationary limit reaches values around $v_{s}$. Hence, in the vicinity of
$v_{s}$, we linearize the dependence on $v$, which is $1/v^{2}$, to $1/v$
and use for $v_{s}$ the expression given by eqn. (\ref{eq:set}) for the
SET model, or by eqn. (\ref{eq:xi}) for $1< \xi <2$. This approximation
allows us to use eqn. (\ref{eq:gvconstant}) to determine $S/\gamma$
provided we can obtain $v_{s}$ and $g_{v}$ from our own experiments.

\subsection{Parameter Estimation}
\label{mle}

\begin{figure}[htbp]
  \begin{center}
    \includegraphics[width=\textwidth]{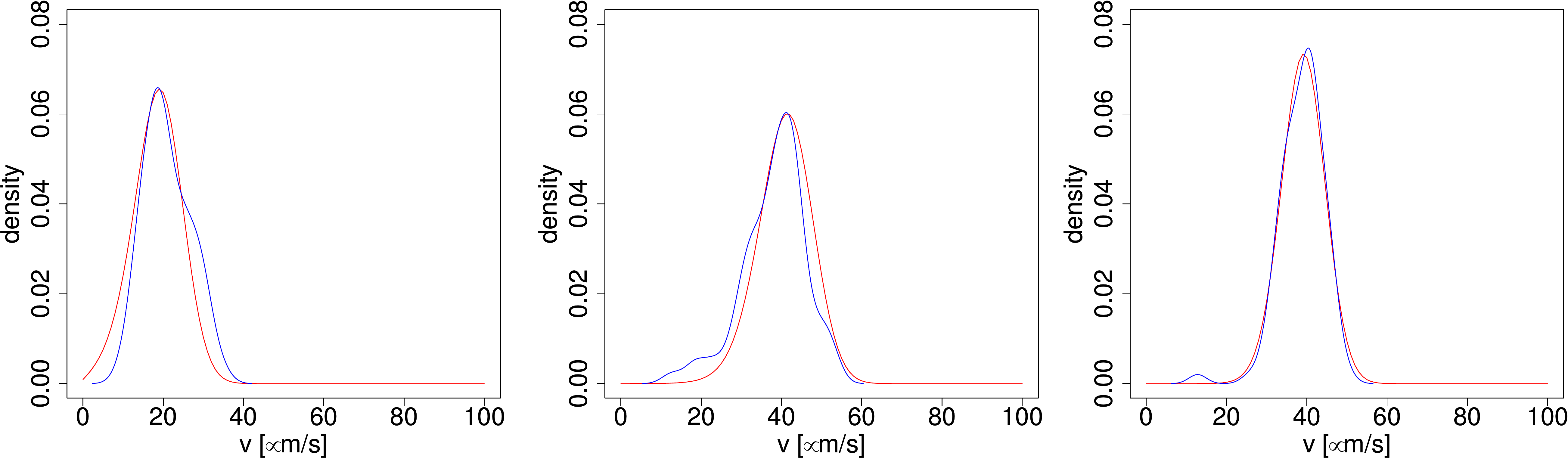}\\
    \includegraphics[width=\textwidth]{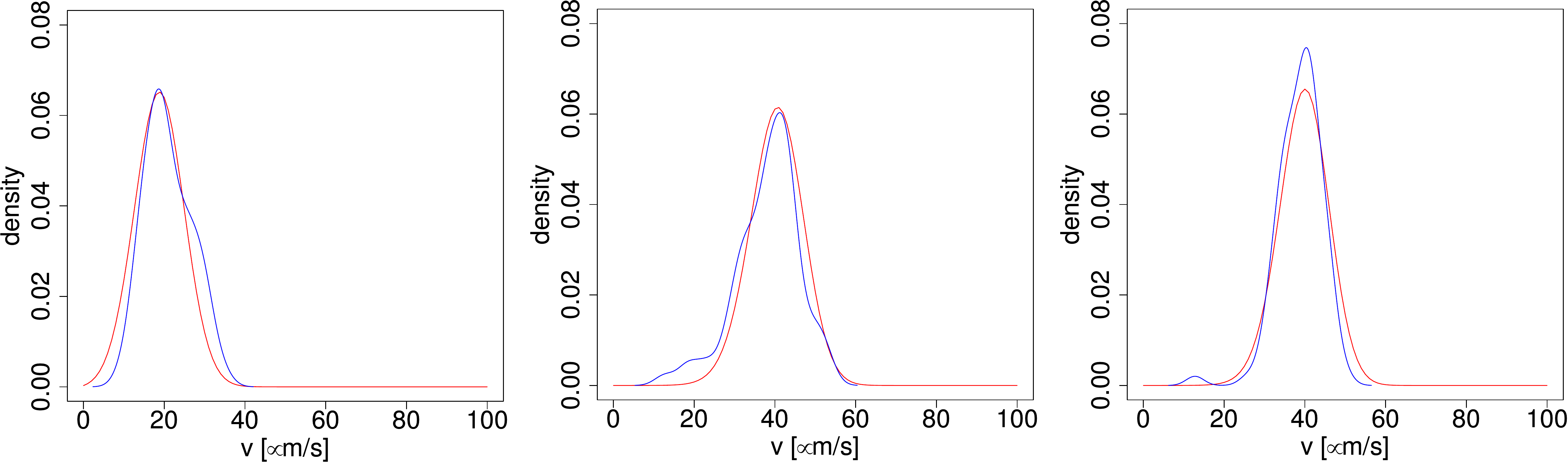}
  \end{center}
  \caption{(blue curves) Estimated densities $p_{a}^{\mathrm{exp}}(v)$
    obtained from experimental measurements of the speed (absolute
    velocity) $v$ $[\mu$m/s] at time $t_{\mathrm{end}}$. (red curves)
    Calculated densities $p_{a}(v)$, eqn. (\ref{pa0}) using the
    parameters of the MLE, Table \ref{ExpParm2}. (top row) SET model with
    $\xi=2$, (bottom row) $1<\xi<2$. (left figures) medium 0, (middle
    figures) medium 1, (right figures) medium 2.}\label{fits}
\end{figure}

With these considerations, we have all ingredients together to compare
our experimental data with the theoretical predictions. In detail. we
proceed as follows: 
\begin{enumerate}
\item For the three media described above (Sect. \ref{sec:setup}), we
  calculate the absolute velocities $v_{aj}^{k }$ of each cell $k$
  tracked during the time step $j$. For each medium, two measurements
  were taken: (a) initially, time $t_{0}$, (b) after sufficiently long
  time, about one hour, time $t_{end}$ (see also Fig. \ref{aveldis} for
  the two-dimensional velocity distributions).  From these samples, which
  contain about 1.500 data points each, we calculate the densities
  $p_{a}^{\mathrm{exp}}(v)$ using the Sheather-Jones method of selecting
  a smoothing parameter for density estimation \citep{sheather1991rdb} in
  R. The results are shown by the \emph{blue curves} in Fig. \ref{fits}
  for $t_{\mathrm{end}}$.
\item From the 6 different density plots, we calculate the maximum
  $v_{s}$ of the experimental speed distribution
  $p_{a}^{\mathrm{exp}}(v)$. The results are given in Appendix C: Table
  \ref{tab:speed}. For $t_{\mathrm{end}}$, we also calculate the speed
  autocorrelation function $g_{v}$, which is shown in Table
  \ref{tab:speed} as well.

\item Eventually, we apply the maximum-likelihood estimation (MLE) to
  find out, which of the still undertermined parameters fit best the
  experimental densities $p_{a}^{\mathrm{exp}}(v)$.  The parameter
  details are presented in Appendix C: Table \ref{ExpParm2}, while the
  resulting density plots for the hypotheses $\xi=2$ and $1<\xi<2$ are
  shown by the \emph{red curves} in Fig. \ref{fits}, which are to be
  compared with the experimental findings (blue curves).
\end{enumerate}

In the following, we further discuss these findings.  In Table
\ref{tab:speed}, one notes slight differences in $v_{s}(t_{0})$ for the
different media. This indicates that at time $t_{0}$, right after being
put into the medium, the bacterial cells already started to adjust to the
nutrient concentration in the media. The higher the concentration, the
higher $v_{s}$. This can be also confirmed at $t_{\mathrm{end}}$, after
about 80 min. The differences between medium 1 and 2 (middle and high
concentration) is rather small both for at $t_{0}$ and
$t_{\mathrm{end}}$, indicating that there seems to be a saturation in
converting internal into kinetic energy. This saturation can be caused by
intracellular processes (e.g. number of receptor proteins available), but
is not further discussed here. Noticable, in medium 0 (no nutrients),
$v_{s}$ \emph{drops down} considerably compared to the initial value,
which also indicates that the bacterial cells respond to the available
energy by adjusting their speed.

We further find that the speed autocorrelation function $g_{v}$ returns
comparable values for all three media ($0.92-0.98$) which are much larger
than for granulocytes, because we have much smaller and more motile
cells. The estimated standard deviations $\sigma=\sqrt{S/\gamma}$ range
between $5.49 \mu$m/s (medium 0) and $5.83 \mu$m/s (medium 2) and are
quite similar for all media, because the temperature and the viscosity of
the media are kept as constant as possible. 

The actual width of the speed distribution, however, does not just depend
on $S/\gamma$ but also on the parameters of the internal energy depot and
is therefore larger than $\sigma$. In order to calculate the parameter
values that maximize the likelihood, we restrict ourselves to reduced
parameters. Keeping in mind that $\sigma^{2}=S/\gamma$ is given by
eqn. (\ref{eq:gvconstant}) and the control parameter is defined as
$Q_{2}=(q_{0}d_{2})/(\gamma m c)$, we can rewrite the leading terms in
eqn. (\ref{pa0}) as:
\begin{equation}
  \label{eq:leading}
  \left( 1+ \frac{d_2 {v}^\xi}{c} \right)^{\frac{q_0}{\xi m S}}
  \exp \left( - \frac{\gamma}{2S} v^2 \right ) =
 \left( 1+ \frac{d_2}{c} {v}^\xi
  \right)^{\frac{Q_{2}}{\xi \sigma^{2}}\frac{c}{d_{2}} }
  \exp \left( - \frac{v^{2}}{2\sigma^{2}} \right )
\end{equation}
which reduces the number of parameters to be determined to $(d_{2}/c)$
and $Q_{2}$, while $\sigma^{2}$ is given by the experiments. $\xi$ on the
other hand is either set to 2, in case of the SET model, or used as a
free parameter.  Given the observations $v_1, v_2, ..., v_n$ the MLE then
determines for which values $\Theta$ of these parameters the likelihood
function $L(v_1, v_2, ..., v_n, \Theta)$ is maximised, i.e. what are the
most likely model parameters that fit the experimental data best,
conditional on the model used.

In order to fully appreciate the MLE, we have to notice that no further
``logical'' assumption are made, i.e., each for experimental distribution
the MLE returns that set of parameter values that fits \emph{this}
particular distribution best. Precisely, we receive most likely a
different set of values for each of the given distributions. To put it
the other way round, from all the observed distributions we can obtain a
range of parameter values that is compatible with the experimental
findings, rather than a precise value that is met by all observations.

With this in mind, we can interpret Table \ref{ExpParm2} as follows: For
both hypotheses, the SET model with $\xi=2$ and the general model with
$1< \xi <2$, we find from the observations a ``reasonable'' (but
different) set of parameters that supports these hypotheses. This is also
confirmed by the fits shows in Figure \ref{fits}. I.e., we cannot reject
one of these hypotheses as they both match the experimental
findings. However, we did not observe a subciritical take-up of energy
for the SET model, as we did not find values $Q_{2}<1$ for the given
observations. The latter conclusion needs a further explanation: In
medium 0, we made no nutrients available, so $q_{0}$ and $Q_{2}$ should
both be zero. However, the bacteria were grown in a medium that contained
nutrients and thus, at time $t_{0}$, started their motion with a filled
energy depot $e^{\mathrm{st}}$, eqn. (\ref{eq:estat}), that changes over
time only rather slowly as $v_{s}$ is adjusted. Hence, the value $Q_{2}$
for medium 0 at time $t_{\mathrm{end}}$ reflects the value of the
internal energy depot at time $t_{\mathrm{end}}$. As the observations in
Figures \ref{aveldis}, \ref{fits} and Table \ref{tab:speed} show, even
after a long time the bacteria still have energy enough to move with a
non-trivial speed $v_{s}$ despite the fact that no nutrients are
provided. But there is a clear trend toward slowing down as the values
indicate.

As second interesting observation regards the decrease of the $d_{2}/c$
values with increasing nutrient concentration (comparing medium 1 and 2),
for both the SET and the $\xi$ model. The ratio $M=d_{2}v_{s}^{2}/c$
reflects the proportion of energy bacteria spend on the two different
processes, active motion and metabolism.  If more energy becomes
available (from medium 1 to 2) , this does not necessarily lead to a
speed-up -- the speed was kept almost constant, but the additional energy
is likely spent on metabolism (and growth). Hence $M$ \emph{decreases}
from 1.50 to 0.62 for the SET model, and from 0.74 to 0.23 for the $\xi$
model, while the take-up of energy $q_{0}/\gamma_{0}$ has
\emph{increased} from 2500 to 4140 for the SET model, and from 5230 to
6800 for the $\xi$ model. So, in conclusion, our model suggests that
bacteria indeed take up more energy from the environment if more
nutrients are provided, but the ratio spent on active motion is
decreased, while the ratio spent on metabolism is increased.

\section{Conclusions}
\label{sec:concl}

The aim of this paper is twofold: (i) we investigate to what extent a
theoretical model of active motion, namely that of ``active Brownian
particles'', is compatible with experimental findings from bacterial cell
motility, (ii) we test the impact of available energy in the environment,
varied by the nutrient concentration in the medium, on the speed
distribution of bacterial cells.

For the discussion one has to keep in mind that our results have been
obtained for a particular strain of \emph{Salmonella typhimurium} (see
Sect. 3.1. and Appendix A) and cannot easily generalized to other
bacterial cells. This is because of the rather complex cellular motion of
bacteria which, in many cases, is comprised of tumbling and nontumbling
phases (see Sect. 3.1). Hence, for the motion of other types of bacteria
we refer to the extensive literature \citep{liao2007pvc, sherwood2003abr,
  berg2003rmb, magariyama2001bss, berry1999bfm, levin1998ois,
  mitchell1991ics, lowe1987rrf, greenberg1977mfb}.

Comparing the experiments with moving \emph{S. typhimurium} and the
theoretical model, we demonstrated that the measured speed distribution
can indeed be matched by the analytical prediction. This holds for both
tested hypothesis, (a) the SET model with $\xi=2$ and (b) the $\xi$
model, which allows an adjustment of the exponent, $1< \xi <2$. We
further confirmed that, under the give experimental conditions, bacteria
move in the supercritical regime, $Q_{2}>1$.  However, we were not able
to observe as subcritical behavior with $Q_{2}<1$ as predicted for the
SET model \citep{ebeling1999abp,condat2002nem}. This can be probably
explained by the fact that even at the end of our experiments bacterial
cells had still enough internal energy available from their growing
period, to move with a nontrivial speed. But we could notice a
considerable slowing down of 25\
with no additional nutrients.

Regarding the impact of available nutrients, we found that the speed of
bacteria did not increase in proportion to it. Instead, we observed a
more or less constant speed, even if the nutrient concentration was
doubled. From calculating the model parameters for both cases, we
conjecture that indeed more energy was taken up by the cells, but this
was used for other internal processes such as metabolism and growth.

In conclusion, the experiments carried out with bacterial cells moving in
media of three different nutrient concentrations could confirm the
theoretical predictions and thus indirectly also support the assumptions
made for our model of active Brownian particles. However, particular
details of the choice of parameters cannot be fully resolved by our
experiments -- which is not very surprising. This regards for example the
``correct'' value of the exponent $\xi$ for the speed, $v^{\xi}$. Our
findings support values between 1.67 and 2.0 if a take-up of energy from
the medium was possible.  The differences between the two assumptions are
not so much in the values of $\xi$ but in the theoretical
consequences. In the case of the SET model, there is a clear bifurcation
which allows to distinguish between subcritical (Brownian motion like)
behavior, and supercritical behavior characterized by a directed
motion. It would still be interesting to find microorganisms for which
these regimes could be determined. Our experiments had to restrict to the
conditions explained above and therefore do not support this
distinction.

\section*{Appendix A: Details of Bacterial Probes}

The chemotaxis-deficient Salmonella strain M935 (SL1344; cheY::Tn10,
\citep{stecher2008mas, winnen2008hep}) was grown under mild aeration for
12 h/37$^{\circ}$C in LB medium containing 0.3 M NaCL. The culture was
diluted 1:20 into fresh medium and grown for another 4h /
37$^\circ$C. Bacteria were pelleted by centrifugation (8500 rpm, 5 min,
4$^\circ$C) and resuspended in phosphate-buffered saline (PBS). The
sample was centrifuged again and the bacteria were resuspended either in
PBS (probe A) or LB medium (probe B). From these two probes, samples for
live imaging of the bacteria were prepared as follows:

\begin{enumerate}
\item probe A was diluted 1:50 in PBS containing 5\
  (BSA) and transferred to a glass-bottom dish for imaging

\item a glass-bottom dish was rinsed with PBS/5\
  of bacteria to the glass surface. PBS/BSA was removed and a 1:50
  dilution of probe B in LB was added to the dish for imaging

\item probe B was diluted 1:50 in a 1:1 mixture of PBS/5\
  (final BSA concentration: 2.5\

\end{enumerate}
All solutions for dilution of probes A and B were prewarmed to
37$^\circ$C.

\section*{Appendix B: Details of Data Evaluation}

For time lapse microscopy the different samples were mounted onto a
heated specimen holder (37$^\circ$C) on a \textrm{Zeiss Asiovert 200m}
inverted microscope. Time series of phase contrast images were recorded
using a Plan Neofluoar 20x (NA 0.5) objective at a rate of ca. 20 images
per second.

Motile bacteria were tracked using Particle tracker software
\citep{sbalzarini2005fpt} as plugin on the pure Java image processing
program ImageJ \citep{abramoff2004ipi}. Only trajectories
appearing over more than 30 time frames were considered. Selected
trajectories were manually verified for correct tracking (Wrong tracking
occurred in cases of crossings between identified bacteria and was
removed from the trajectories). These criteria for track evaluation were
equally applied on all detected trajectories. About 3 hours of eye
selection are necessary to obtain about fifteen to twenty tracks. The
evaluation of the data was carried out by an R-script written
by the authors.

\clearpage 
\section*{Appendix C: Details of Parameters Obtained}

\begin{table}[!h]
  \centering
  \begin{tabular}{|c|c|c|c|}
    \hline 
    Medium    & $v_{s}(t_{0})$ [$\mu$m/s] & $v_{s}(t_{\mathrm{end}})$
    [$\mu$m/s] & $g_{v}(t_{\mathrm{end}})$ \\ \hline \hline
    0  & 24.2 & 19.8 &  0.920 \\ \hline  
    1  & 27.8  &  38.8 & 0.979 \\ 
    2  & 29.3  &  39.5 & 0.981 \\ \hline  
  \end{tabular}
  \caption{Maxima $v_s$ (10$^{-6}$ m/s) of the experimental speed distribution
    $p_{a}^{\mathrm{exp}}(v)$ taken initially ($t_{0}$) and after long
    time ($t_{\mathrm{end}}$) for 3 different media (see
    Sect. \ref{sec:setup}). $g_{v}(t_{\mathrm{end}})$ gives the value of the speed
    autocorrelation function measured experimentally at time
    $t_{\mathrm{end}}$.}
  \label{tab:speed}
\end{table}

\begin{table}[!h]
\begin{center}
  \begin{tabular}{|c||c|c||c|c|c|}
    \hline Medium & ${d_{2}}/{c}$
    [(s$/(\mu$m$)^{2}$] & $Q_{2}$ & $^{\star}{d_{2}}/{c}^{\star}$
    [(s$/(\mu$m$)^{2}$] & $^{\star}Q_{2}^{\star}$ & $\xi$ \\ \hline
    \hline 0 & $2.3\cdot 10^{-3}$ & 1.75 & $33.3\cdot 10^{-4}$ & 8.50 &
    1.52 \\ \hline 1 & $1.0\cdot 10^{-3}$ & 2.63 &
    $\;\; 9.5\cdot 10^{-4}$  & 4.97 & 1.82 \\
    2 & $0.4\cdot 10^{-3}$ & 1.70 & $\; \; 5.1\cdot 10^{-4}$ & 3.47 &
    1.67 \\ \hline
  \end{tabular}
  \caption{Parameter values of $p_{a}(v)$, eqn. (\ref{pa0}), as estimated
    by MLE for the SET model, $\xi=2$ (no-starred values) and for the
    general model, $1<\xi<2$ (starred values) at time
    $t_{\mathrm{end}}$. }
\label{ExpParm2}
\end{center}
\end{table}

\subsection*{Acknowledgment}

The authors are deeply indebted to Wolf-Dietrich Hardt for providing
access to, and use of, his laboratory at the Institute of Microbiology of
ETH Zurich, where V.G. could carry out the experiments. We further
gratefully acknowledge scientific discussions with Howard C. Berg,
Wolf-Dietrich Hardt and Markus C. Schlumberger.

\end{document}